\documentstyle[twoside,11pt,epsfig]{article}
\catcode`\@=11
\long\def\@makefntext#1{
\protect\noindent \hbox to 3.2pt {\hskip-.9pt  
$^{{\tenrm\@thefnmark}}$\hfil}#1\hfill}                 
 
\def\thefootnote{\fnsymbol{footnote}}
\def\@makefnmark{\hbox to 0pt{$^{\@thefnmark}$\hss}}    
 
\def\ps@myheadings{\let\@mkboth\@gobbletwo
\def\@oddhead{\hbox{}
\rightmark\hfil\tenrm\thepage}  
\def\@oddfoot{}\def\@evenhead{\tenrm\thepage\hfil
\leftmark\hbox{}}\def\@evenfoot{} 
\def\sectionmark##1{}\def\subsectionmark##1{}}

\renewcommand{\thefootnote}{\fnsymbol{footnote}}

\newcounter{sectionc}\newcounter{subsectionc}\newcounter{subsubsectionc}
\renewcommand{\section}[1] {\vspace{25pt}\addtocounter{sectionc}{1}
\setcounter{subsectionc}{0}\setcounter{subsubsectionc}{0}\noindent 
			{\twelvebf\thesectionc.\kern0.35cm #1}\par\vspace{8pt}}
\renewcommand{\subsection}[1] {\vspace{25pt}\addtocounter{subsectionc}{1}
			\setcounter{subsubsectionc}{0}\noindent
			{\twelvebf\thesectionc.\thesubsectionc\kern0.35cm #1}\par 
			\vspace{8pt}}
\renewcommand{\subsubsection}[1] {\vspace{25pt}\addtocounter{subsubsectionc}{1}
			\noindent 
			{\twelverm\thesectionc.\thesubsectionc.\thesubsubsectionc\kern0.35cm
			{\kern1pt\twelveit #1}}\par\vspace{8pt}}

\newcommand{\smalllineskip}{\baselineskip=11pt}

\def\ninecirc{ 
\begin{picture}(0,0)
\put(4.4,1.8){\circle{7.45}} 
\end{picture}}
\def\ninecopyright{\ninecirc\kern2.75pt\hbox{\eightrm c}}
 
\newcommand{\copyrightheading}[1] 
			{\vspace*{-1cm}\baselineskip=11pt{\flushleft
			{\ninerm Fractals, #1}\\
			{\ninerm $\ninecopyright$\,\,\, World Scientific 
			Publishing Company}\\
			}}

\def\abstracts#1#2#3{{ 
					\centering{\begin{minipage}{5.0in}\tenrm\baselineskip=12pt
					\centerline{\twelvebf Abstract}
					\vspace{5pt}
					\parindent=0pc #1\par 
					\parindent=1pc #2\par
					\parindent=1pc #3
					\end{minipage}}\par}} 

\def\ARTICLES{\kern6.15cm\hbox{${\vcenter{\vbox{
	\hrule height 0.4pt width 4.562truein       
	\hbox{\vrule width 0.4pt                    
height 0.6truein                            
\raise0.565cm\hbox{\kern1pc
	\seventeenbf Articles}}
	\hrule height 0.4pt width 4.562truein}}}$}}  

\renewenvironment{thebibliography}[1]
	{\frenchspacing
		\tenrm\baselineskip=12pt
	\begin{list}{\arabic{enumi}.} 
{\usecounter{enumi}\setlength{\parsep}{0pt}
		\setlength{\leftmargin 12.7pt}{\rightmargin 0pt} 
	\setlength{\itemsep}{0pt} \settowidth 
{\labelwidth}{#1.}\sloppy}}{\end{list}}

\newcounter{itemlistc}
\newcounter{romanlistc}
\newcounter{alphlistc}
\newcounter{arabiclistc}

\newcommand{\fcaption}[1]{
	\refstepcounter{figure}
\setbox\@tempboxa = \hbox{\footnotesize{\bf Fig.~\thefigure\phantom{00}}#1} 
\ifdim \wd\@tempboxa > 6in
			{\begin{center}
\parbox{6in}{\footnotesize\smalllineskip{\bf Fig.~\thefigure\phantom{00}}#1} 
\end{center}}
	\else
					{\begin{center}
					{\footnotesize{\bf Fig.~\thefigure\phantom{00}}#1}
						\end{center}}
	\fi}

\newcommand{\tcaption}[1]{
	\refstepcounter{table}
\setbox\@tempboxa = \hbox{\footnotesize\bf Table~\thetable\phantom{00}#1} 
\ifdim \wd\@tempboxa > 6in
			{\begin{center}
\parbox{6in}{\footnotesize\smalllineskip\bf Table~\thetable\phantom{00}#1} 
\end{center}}
	\else
					{\begin{center}
					{\footnotesize\bf Table~\thetable\phantom{00}#1}
						\end{center}}
	\fi}

\def\@citex[#1]#2{\if@filesw\immediate\write\@auxout  
{\string\citation{#2}}\fi
\def\@citea{}\@cite{\@for\@citeb:=#2\do 
{\@citea\def\@citea{,}\@ifundefined
	{b@\@citeb}{{\bf ?}\@warning
	{Citation `\@citeb' on page \thepage \space undefined}}
	{\csname b@\@citeb\endcsname}}}{#1}}

\newif\if@cghi
\def\cite{\@cghitrue\@ifnextchar [{\@tempswatrue 
\@citex}{\@tempswafalse\@citex[]}}
\def\citelow{\@cghifalse\@ifnextchar [{\@tempswatrue 
\@citex}{\@tempswafalse\@citex[]}}
\def\@cite#1#2{{$\null^{#1}$\if@tempswa\typeout
	{IJCGA warning: optional citation argument 
	ignored: `#2'} \fi}}

\def\fnt#1#2{\footnotetext{\kern-.3em
	{$^{\mbox{\sevenrm #1}}$}{#2}}}

\def\runninghead#1#2{\protect\pagestyle{myheadings} 
\markboth{\protect\nineit\,\,\,\,\,#1\hfill}
{\hfill\protect\nineit #2\,\,\,\,\,}}
\font\seventeenbf=cmbx10      scaled\magstep3 

\font\twelverm=cmr10      scaled\magstep1
\font\twelveit=cmti10     scaled\magstep1
\font\twelvebf=cmbx10     scaled\magstep1

\font\tenrm=cmr10
\font\tenit=cmti10
\font\tenbf=cmbx10

\font\ninerm=cmr9
\font\nineit=cmti9

\font\eightrm=cmr8

\font\sevenrm=cmr7

\catcode`@=11                                   
\def\ps@plain{\let\@mkboth\@gobbletwo 
\def\@oddhead{}\def\@oddfoot{\ninerm\hfil\thepage
\hfil}\def\@evenhead{}\let\@evenfoot\@oddfoot}

\def\ps@myheadings{\let\@mkboth\@gobbletwo      
\def\@oddhead{\hbox{} 
\rightmark\hfil\ninerm\thepage}  
\def\@oddfoot{}\def\@evenhead{\ninerm\thepage\hfil
\leftmark\hbox{}}\def\@evenfoot{} 
\def\sectionmark##1{}\def\subsectionmark##1{}}
\catcode`@=12 

\begin{document} 

\runninghead{Nonlinear Relaxation in Population Dynamics}
{Nonlinear Relaxation in Population Dynamics}

\renewcommand{\thefootnote}{\fnsymbol{footnote}}     
\thispagestyle{plain}
\setcounter{page}{1}

\copyrightheading{Vol.~0, No.~0 (0000)}

\vspace{6pc}
\leftline{\phantom{\ARTICLES}\hfill}

\vspace{3pc}
\leftline{\hskip-0.1cm\vbox{\hrule width6.99truein height0.15cm}\hfill}

\vspace{2pc} 
\centerline{\seventeenbf NONLINEAR RELAXATION}
\baselineskip=20pt 
\centerline{\seventeenbf IN}
\centerline{\seventeenbf POPULATION DYNAMICS} 
\vspace{0.27truein}
\centerline{MARKUS A. CIRONE}
\baselineskip=12.5pt
\centerline{\it INFM, Unit\'a di Palermo, and Dipartimento di Fisica e
Tecnologie Relative} 
\centerline{\it Universit\'a di Palermo, Viale delle Scienze, Palermo, I-90128,
Italy}  \vspace{0.08truein} \centerline{FERDINANDO DE PASQUALE}
\baselineskip=12.5pt
\centerline{\it INFM, Unit\'a di Roma, and Dipartimento di Fisica, 
 Universit\'a di Roma}
\centerline{\it "La Sapienza", P.le A. Moro 2, Roma, I-00185, Italy}
\vspace{0.08truein} \centerline{BERNARDO SPAGNOLO}
\baselineskip=12.5pt 
\centerline{\it INFM, Unit\'a di Palermo, and Dipartimento di Fisica e
Tecnologie Relative} 
\centerline{\it  Universit\'a di Palermo, Viale delle Scienze, Palermo, I-90128,
Italy} 
  \vspace{0.36truein}

\abstracts{We analyze the nonlinear relaxation 
of a complex ecosystem composed of many interacting species. 
The ecological system is described by generalized Lotka-Volterra 
equations with a 
multiplicative noise. The transient dynamics is  studied 
in the framework of the mean field theory and with random  
interaction between the species. We focus on the statistical 
properties of the asymptotic behaviour of the time integral 
of the  {\em i}-th population and on the distribution of the 
population  and of the local field.}{}{}

\vspace{0.78truein} 
\baselineskip=14.5pt
\section{INTRODUCTION}

\noindent
Systems of interacting biological species evolve through a dynamical 
complex process that can be conveniently described, within 
relatively short time scales, by generalized Lotka-Volterra equations
\cite{SviLog}. The nonlinearity of these equations complicates their
analytical investigation, especially in the case of a great number of
interacting species.  


\noindent
 Nevertheless some analytical
approximation for the mean field interaction between the species
 as well as numerical simulations give some insight into the
behaviour of complex ecosystems \cite{CiudeSpa,deSpa,CirdeSpa}. 
Basic elements of a Lotka-Volterra model are the growth parameter 
and the interaction parameter. For a large number of interacting 
species, it is reasonable, as a phenomenological approach, to 
choose these parameters at random from given probability distributions. 
Within this type of representation, the dynamics of coevolving 
species can be characterized by statistical properties over different
realizations of parameter sets. Though the generalized
Lotka-Volterra model has been explored in detail \cite{SviLog}, 
it seems that a full characterization, either deterministic or 
statistical, of the conditions under which a population   
extinguishes or survives in the competition process 
has not been achieved \cite{Abram,AbramZa}.

In this paper we analyze the role of the noise on the transient 
dynamics of the ecosystem of many interacting species in the 
presence of an absorbing barrier, i. e. extinction of the species. 
Two type of interaction between the species are considered: 
(a) mean field interaction, and (b) random interaction.
	We focus on the asymptotic behaviour of the time integral 
of the {\em i}-th population and on the distributions of the population 
and of the local field, which is the total interaction of all species 
on the {\em i}-th population.

By introducing an approximation for the time integral of the average
species concentration $M(t)$ we obtain analytical results for the 
transient behaviour and the asymptotic statistical properties 
of the time average  of the {\em i}-th population. We find
that for a very large number of interacting species 
the statistical properties of
the time average of the {\em i}-th population process are determined 
asymptotically from the statistical properties of the Wiener process. 
At the critical point and around the stability-instability  
transition, for mean field interaction, the system goes 
from a purely  long time tail behaviour (namely 
$M(t) \sim \sqrt{t}$ to a new long time modified regime 
$M(t) \sim \sqrt{t} e^{\sqrt{t}}$. 
Specifically for random interaction we find that the
local field  and the cavity field \cite{MePaVir}, which is 
the total interaction
of all species on the {\em i}-th population  when this population is
absent,  are
different in absence of external noise while overlap quite well 
in the presence
of the noise. This behaviour, which is very different from 
the analogue 
spin glasses problem in statistical mechanics, is reminiscent 
of a phase transition phenomenon. It suggest that, becuase all
population are positive and can grow during the dynamical process
of the ecosystem,
each population play an important role on the total interaction
between the species.

The paper is organized as follows. In the next section 
we describe the model. The mean field and the random interactions
are considered in section $3$ and $4$. Our results are
discussed in section 5.

\section{THE MODEL}

\noindent
We consider an N-species
generalization of the usual Lotka-Volterra system with a 
Malthus-Verhulst modelization of the self regulation processes 
for a fully connected ecological network. Therefore the 
Ito stochastic differential equation describing the dynamical 
evolution of the ecosystem is 

							\begin{equation} 
		d n_i(t) = \left[\left(\gamma + \frac{\epsilon}{2} \right) - 
n_i(t) + \sum_{j\neq i} J_{ij}n_j(t) \right]n_i(t) dt + 
\sqrt{\epsilon} n_i(t) dw_i\mbox{,} \,\,\,\,\,\, i = 1,...,N
							\label{langevin}
							\end{equation}
where $n_i(t) \geq 0$ is the number of elements of the {\em i}-th
species.  In Eq.(\ref{langevin}) $\gamma$ is the growth parameter, 
the interaction matrix $J_{ij}$ modelizes 
the interaction between different species ($i\neq j$) and 
$w_i$ is the Wiener process whose increment $dw_i$ satisfies the usual
statistical properties
						
						\begin{equation} <dw_i(t)>\mbox{} =\mbox{}0\mbox{;} \;\;\;\; 
<dw_i(t)dw_j(t^{\prime})> \mbox{}=
\mbox{}\delta_{ij}\delta(t-t^{\prime}) dt.
						\end{equation}
Our ecosystem is composed of $N = 1000$ species. We consider all species
equivalent so that the characteristic  parameters of the ecosystem are
independent of the species. The random interaction with 
the environment (climate,
disease,etc...) is taken into account by introducing a 
multiplicative noise in the Eq.(\ref{langevin}).  
 The solution of the dynamical 
equation Eq.(\ref{langevin}) is given by

						\begin{equation} n_i(t) = \frac{n_i(0)exp\left[
\delta t +\sqrt{\epsilon}
w_i(t) + \int_{0}^{t} dt^{\prime}\sum_{j\neq i}J_{ij}n_j(t^{\prime}) \right]}
{1+\gamma n_i(0) \int_{0}^{t}dt^{\prime}
exp\left[\delta t^{\prime} +\sqrt{\epsilon}
w_i(t^{\prime}) + \int_{0}^{t^{\prime}} dt^{\prime\prime}
\sum_{j\neq i}J_{ij}n_j(t^{\prime\prime})\right]}.
						\label{sol langevin}
						\end{equation}
We consider two
different types of interaction between the species:
(a) a mean field approximation with a symbiotic interaction
between the species; (b) a random interaction
between the species with different types of mutual
interactions: competitive, symbiotic
and prey-predator relationship.

\section{MEAN FIELD APPROXIMATION}

\noindent 
We consider a mean field symbiotic interaction between the species.
As a consequence the growth
parameter is proportional to the  average species concentration 

	\begin{equation} 
\sum_{j\neq i}J_{ij}n_j(t) = \frac{J}{N}\sum_j n_j(t) = J m(t),
							\label{Jm}
							\end{equation}
and the stochastic differential equation Eq.(\ref{langevin}) becomes

	\begin{equation} d n_i = \left[\left(J m  + \gamma +
\frac{\epsilon}{2}\right) n_i - n^2_i\right]dt +
\sqrt{\epsilon}n_i dw_i.
							\label{langevin meanfield}
							\end{equation}
In the limit of a large number of interacting
species the stochastic evolution of the system is given by the following
integral equation

					\begin{equation} M(t) =
\frac{1}{N}\sum_i\ln\left(1+ n_i(0) \int_{0}^{t} dt^{\prime}
e^{J M(t^{\prime})+\gamma t^{\prime}+\sqrt{\epsilon} w_i(t^{\prime})}
\right) ,
       \label{integral eq}
							\end{equation}
where 

							\begin{equation} M(t) = \frac{1}{N}\sum_i\int_{0}^{t} dt^{\prime}
n_i(t^{\prime}) = \int_{0}^{t} dt^{\prime}m(t^{\prime}).
       \label{M(t)}
							\end{equation}
is the time integral of the site population concentration average.
We introduce an approximation of Eq.(\ref{integral eq}) which greatly
simplifies the noise affected evolution of the system and allows us 
to obtain
analytical results for the population dynamics. We note that in this
approximation the noise influence is taken into account in a 
nonperturbative
way, and that the statistical properties of the time average process M(t) are
determined asymptotically from the statistical properties of the process
$w_{max}(t)$ = $\sup_{0<t^{\prime}<t} w(t^{\prime})$,
where $w$ is the Wiener process.
 Starting from the following approximated integral
equation for $M(t)$

					\begin{equation} M(t) \simeq 
\frac{1}{N}\sum_i\ln\left(1 + 
n_i(0) e^{\sqrt{\epsilon} w_{max_i}}
\int_{0}^{t} dt^{\prime}
e^{J M(t^{\prime})+\gamma t^{\prime}}\right)
				  \label{M(t) approx}
						\end{equation}
it is possible to analyze the role of the noise on the stability-instability
transition in three different regimes of the nonlinear relaxation of the 
system: (i) towards the equilibrium population ($\gamma>0$), (ii) towards the
absorbing barrier ($\gamma<0$), (iii) at the critical point ($\gamma=0$).
Specifically at the critical point we obtain for the time average process
$M(t)/t$ as a dominant asymptotic behaviour in the stability region
(namely when $J < 1$)

						\begin{equation} 
						\frac{M(t)}{t} \simeq \left(
\frac{1}{1-J} \right) \sqrt{\frac{2 \epsilon}{\pi}}\frac{1}{\sqrt{t}},
						\end{equation}
and in the instability region (namely when $J>1$)

							\begin{equation}
								\frac{M(t)}{t} \simeq 
e^{\left<\ln(n_i(0))\right>} \sqrt{\frac{2\pi}{\epsilon}}
\frac{exp\left[\sqrt{\frac{2 \epsilon}{\pi}} \sqrt{t}\right]}{\sqrt{t}}
								\end{equation}
We obtain for the case (i) an explicit expression of the transition 
time $t_c$ as a function of the noise intensity ($\epsilon$), the initial
population distribution ($n_i(0)$) and the parameters of the system
($\gamma,J$)

								\begin{equation} t_c \simeq \frac{1}{\gamma}\left\{\left
[\frac{\epsilon}{2\pi\gamma} + \ln\left(1 + \left(\frac{\gamma}
{(J-1)}\right) e^{-\left<\ln(n_i(0))\right>}
\right)\right]^{1/2} - \sqrt{\frac{\epsilon}{2\pi\gamma}}\right\}^2  
							\label{t_c d>0}
							\end{equation}
For the cases (ii) and (iii) we obtain two implicit expressions 
in terms of exponential and error functions of the same
quantities: $t_c, \epsilon$ and the system paramters\cite{deSpa}.
 The transition
time increases from $\gamma>0$ to $\gamma<0$ according to the following
inequality

			\begin{equation}
(t_c)_{\gamma<0}>(t_c)_{\gamma=0}>(t_c)_{\gamma>0}.
						\label{t_c epsilon­0}
							\end{equation}
This means that when the interaction between the species 
prevails over the resources, the presence of a hostile 
environment ($\gamma<0$) causes a late start of the 
divergence of some population (i.e. the instability).
The noise forces the system to sample more of the available range in the
parameter space and therefore moves the system towards the instability.
The effect of the noise is to make unstable the system earlier than in the
deterministic case ($\epsilon =0$).
If we raise the intensity of the noise and keep fixed the initial
distribution, we obtain the same effect of the enhancement of the variance of
the Gaussian initial distribution of the population for 
moderate values of noise intensity (namely $\epsilon = 0.1$).
For high values of noise intensity (namely $\epsilon = 1$) we strongly 
perturb the population dynamics and because of the presence of an
absorbing barrier we 
obtain quickly the extinction of the
populations.

\section{RANDOM INTERACTION}
\noindent 
The interaction 
between the species is assumed to be random and it 
is described by a random interaction matrix $J_{ij}$, 
whose elements are independently distributed according
to a Gaussian distribution

					\begin{equation}
					 P(J_{ij}) = \frac{1}{\sqrt{2\pi\sigma^{2}_J}} 
exp\left[-\frac{J^{2}_{ij}}{2\sigma^{2}_J}\right] \mbox{,} \,\,\,
\sigma^{2}_J = \frac{J^2}{N}.
						\label{P(J)}
							\end{equation}
where $J$ is the interaction strength and

					\begin{equation}
						<J_{ij}> = 0 \mbox{,} \,\,\,\,   \mbox{} \ \ \ <J_{ij} J_{ji}> = 0.
					\end{equation}
With this asymmetric interaction matrix our ecosystem
contains $50\%$ of prey-predator interactions (namely
$J_{ij}<0$ and $J_{ji}>0$), $25\%$ competitive interactions
($J_{ij}<0$ and $J_{ji}<0$) and $25\%$ symbiotic interactions
($J_{ij}>0$ and $J_{ji}>0$).
The initial values of the populations $n_i(0)$ have also 
Gaussian distribution

					\begin{equation}
					 P(n) = \frac{1}{\sqrt{2\pi\sigma^{2}_n}} 
exp\left[-\frac{(n - <n>)^2}{2\sigma^{2}_n}\right]\mbox{,} \,\,\,\,
\sigma^{2}_n = 0.01\mbox{,}\,\,  \mbox{and} <n> = 1.
						\label{P(n)}
							\end{equation}
The strength of interaction between the species $J$ determines two
differnt dynamical behaviours of the ecosystem.
Above a critical value $J_c$ the system is unstable, 
this means
that at least one of the populations diverges. 
Below the critical
interaction strength, the system is stable and reaches 
asymptotically an
equilibrium state. For our ecosystem this critical value is
approximately $J = 1.1$.The equilibrium values
of the populations  depend both on their initial values 
and on the interaction
matrix.  If we consider a quenched random interaction matrix,
the ecosystem has a great number of equilibrium
configurations, each one with its attraction basin.
For vanishing noise
($\epsilon = 0$), the steady state solutions of 
Eq.(\ref{langevin}) are obtained
by the fixed-point equation

						\begin{equation}
						(\gamma - n_i + h_i) n_i = 0
						\label{fixpoint}
						\end{equation}
where

						\begin{equation}
							 h_i = \sum_j J_{ij} n_j(t)
				\label{local field}
				\end{equation}
is the local field. For a great number of interacting species
we can assume that the local field $h_i$ is Gaussian with zero
mean and variance $\sigma^{2}_{h_i} = <h^{2}_i> = J^2 <n^{2}_i>$

		\begin{equation}
	 P(h_i) = \frac{1}{\sqrt{2\pi\sigma^{2}_{h_i}}} 
exp\left[-\frac{h^{2}_{i}}{2\sigma^{2}_{h_i}}\right] 
						\label{P(hi)}
							\end{equation}
 The solutions
of Eq.(\ref{fixpoint}) are

				\begin{equation}
				n_i = 0\mbox{, i. e. extinction} 
			\end{equation}
and
					\begin{equation}
  n_i = (\gamma + h_i)\Theta(\gamma + h_i)\mbox{,} \,\,\, n_i>0,
					\end{equation} 
where $\Theta$ is the Heaviside unit step function. From this equation
and applying the self consistent condition we can calculate 
the steady state average population and its variance.
Specifically we have

					\begin{eqnarray}
				<n_i> & = & \left<(\gamma 
+ h_i) \Theta(\gamma + h_i)\right> = \nonumber \\
&	= &\frac{1}{\sqrt{2\pi\sigma^{2}_{h_i}}}
\left[\sigma^{2}_{h_i}
exp\left[\frac{\gamma^2}{2 \sigma^{2}_{h_i}}\right] +
\frac{\gamma\sqrt{2 \sigma^{2}_{h_i}\pi}}{2}
\left(1 + erf \left(\frac{\gamma}{\sqrt{2 \sigma^{2}_{h_i}}}
\right)\right)\right],
				\label{average}
				\end{eqnarray}
and

					\begin{eqnarray}
						<n^{2}_i> & = &\left<(\gamma + 
h_i)^2 \Theta^{2}(\gamma + h_i) \right> = \nonumber \\
& = & \left[\left(\frac{\gamma^2 + 
\sigma^{2}_{h_i}}{2} \right) \left(1 + 
erf \left(\frac{\gamma}{\sqrt{2\sigma^{2}_{h_i}}}
\right)\right) + \frac{\gamma}{2}
\sqrt{\frac{2\sigma^{2}_{h_i}}{\pi}} 
exp\left[\frac{\gamma^2}{2 \sigma^{2}_{h_i}}\right]\right].
						\label{variance}
						\end{eqnarray}
For an interaction strength $J = 1$ and 
an intrinsec growth parameter $\gamma = 1$ we obtain:
$<n_i> = 1.4387, <n^{2}_i = 4.514,$ and $\sigma^{2}_{n_i} = 2.44$.
These values are in good agreement with that obtained from
numerical simulation of Eq. (\ref{langevin}).
The choice of this particular value for the interaction strength,
based on a preliminar investigation on the stability-instability
transition of the ecosystem, ensures us that the ecosystem is stable.

\begin{figure}[b!] 
\centerline{\epsfig{file=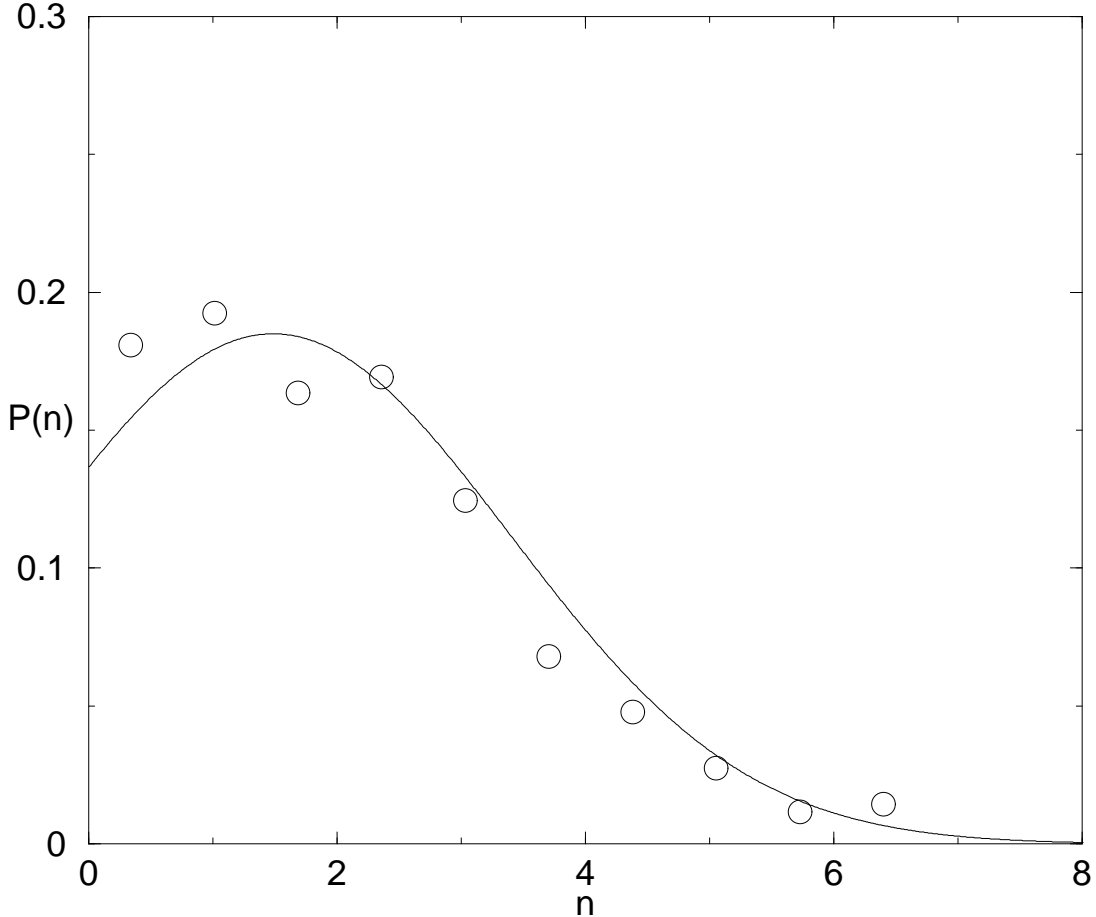,width=5.7in}}
\vspace{10pt}Ê\caption{The stationary probability distribution 
(Eq.(\ref{P(ni)_zeronoise})), without the extinct species,
 in comparison with the histograms arising from
numerical simulations (open circles). The system partameters are:
$N = 1000$ species, $J = 1$ and $\gamma = 1$.}
	\label{P(ni)}
	\end{figure}

\noindent
 The stationary probability
distribution of the populations is the sum of a delta 
function and a truncated
Gaussian

		\begin{equation}
			P(n_i) = n_{e_i} \delta (n_i) + 
\Theta(n_i) \frac{exp\left[-\frac{(n_i - n_{io})^2}
{2 J^2 \sigma^{2}_{n_i}}\right]}
{\sqrt{2 \pi J^2 \sigma^{2}_{n_i}}}.
			\label{P(ni)_zeronoise}
			\end{equation}
In Fig.(1) we report the stationary probability distribution 
of the population densities, without the extinct species,
in comparison with the
computer simulations for systems with N = 1000 species and for 
an interaction strength $J = 1$, and $\gamma = 1$.

 As in the previous case we focus on the statistical properties 
of the time integral of the {\em i}-th population $N_i(t)$

\begin{equation} N_i(t) = \int_{0}^{t} dt^{\prime} n_i(t^{\prime}),
				\label{Ni(t)}
				\end{equation}
in the asymptotic regime. From Eq. (\ref{sol langevin}) we have
 
				\begin{equation} N_i(t) =
\ln\left[1+ n_i(0) \int_{0}^{t} dt^{\prime}
exp\left[\gamma t^{\prime}+\sqrt{\epsilon} w_i(t^{\prime}) + 
\sum_{J \neq i} J_{ij} N_j(t^{\prime})\right]\right] ,
				\label{Ni Integr. eq.}
				\end{equation}
In Eq. (\ref{Ni Integr. eq.}) the term $\sum_j J_{ij}N_j$ 
gives the influence of
other species on the differential growth rate of 
the time integral of the {\em i}-th population and represents 
a local field acting on the
ith population \cite{deSpa,MePaVir}

				\begin{equation} h_i = \sum_j J_{ij} N_j(t) = J \eta_i.
				\label{N_localfield}
				\end{equation}
 We  use the same approximation of the 
Eq.(\ref{M(t) approx}) and, after differentiating,
we get the asymptotic solution of Eq.(\ref{Ni Integr. eq.})

				\begin{equation} N_i(t) \simeq
 \ln\left[ n_i(o) e^{\sqrt{\epsilon} w_{max_i}(t) + 
J \eta_{max_i}(t)} \int^t_0 dt^{\prime} e^{\gamma t^{\prime}}\right]
					\label{Ni g>0}
					\end{equation}
where $w_{max_i}(t) = sup_{0<t^{\prime}<t}w(t^{\prime})$ and
$ \eta_{max_i}(t) = sup_{0<t^{\prime}<t}\eta(t^{\prime})$.
The Eq.(\ref{Ni g>0} is valid for
$\gamma \geq 0$, that is when  the system relaxes towards an
equilibrium population and at the critical point.
Evaluating Eq.(\ref{Ni g>0}) for $\gamma \geq 0$, after
making the ensemble average, we obtain for the time average
of the  {\em i}-th population $\bar{N_i}$

			\begin{equation}
\left<\bar{N_i}\right> \simeq \frac{1}{t}
\left[N_w \sqrt{\epsilon t}
 + \ln t + \left<\ln\left[n_i(o) \right]\right>\right]
\mbox{,} \,\,\, \gamma = 0,
			\label{Ni g=0}
   \end{equation}
and
					\begin{equation}
\left<\bar{N_i}\right> \simeq \frac{1}{t}
\left[N_w \sqrt{\epsilon t}
 + (\gamma + N_{\eta} + \left<\ln\left[\frac{n_i(o)}{\gamma}
\right]\right>\right] \mbox{,} \,\,\, \gamma > 0,
			\label{Ni g>0_eps}
   \end{equation}
where $N_w$ and $N_{\eta}$ are variables with a semi-Gaussian
distribution \cite{CiudeSpa} and $N_{\eta}$ must be 
determined self-consistently from the 
Eq. (\ref{N_localfield}). These asymptotic behaviours are
consistent with those obtained using a 
mean field approximation. We obtain in fact the typical
long time tail behaviour ($t^{-1/2}$) dependence, which
characterize nonlinear relaxation regimes when
$\gamma \geq 0$. Besides
the  numerical results confirm these analytical
asymptotic behaviours of $\bar{N_i}$ \cite{CirdeSpa}.
When the system
relaxes towards the absorbing barrier ($\gamma<0$) we get 
from Eq. (\ref{Ni Integr. eq.})
in the long time regime

				\begin{equation}
\left<\bar{N_i}\right> \simeq \frac{1}{t}
\left[\ln(n_i(0)) +
\ln\left[ \int^t_0 dt^{\prime} e^{\gamma t^{\prime} +
\sqrt{t^{\prime}} w_i(t^{\prime}) + j \eta_i(t^{\prime})} \right]\right].
				\label{Ni g<0}
				\end{equation}
In this case the time average of the {\em i}-th population 
$\left<\bar{N_i}\right>$ is 
a functional of the local field and the Wiener process, and
it depends on the history of these two stochastic
processes.

\begin{figure}[b!] 
\centerline{\epsfig{file=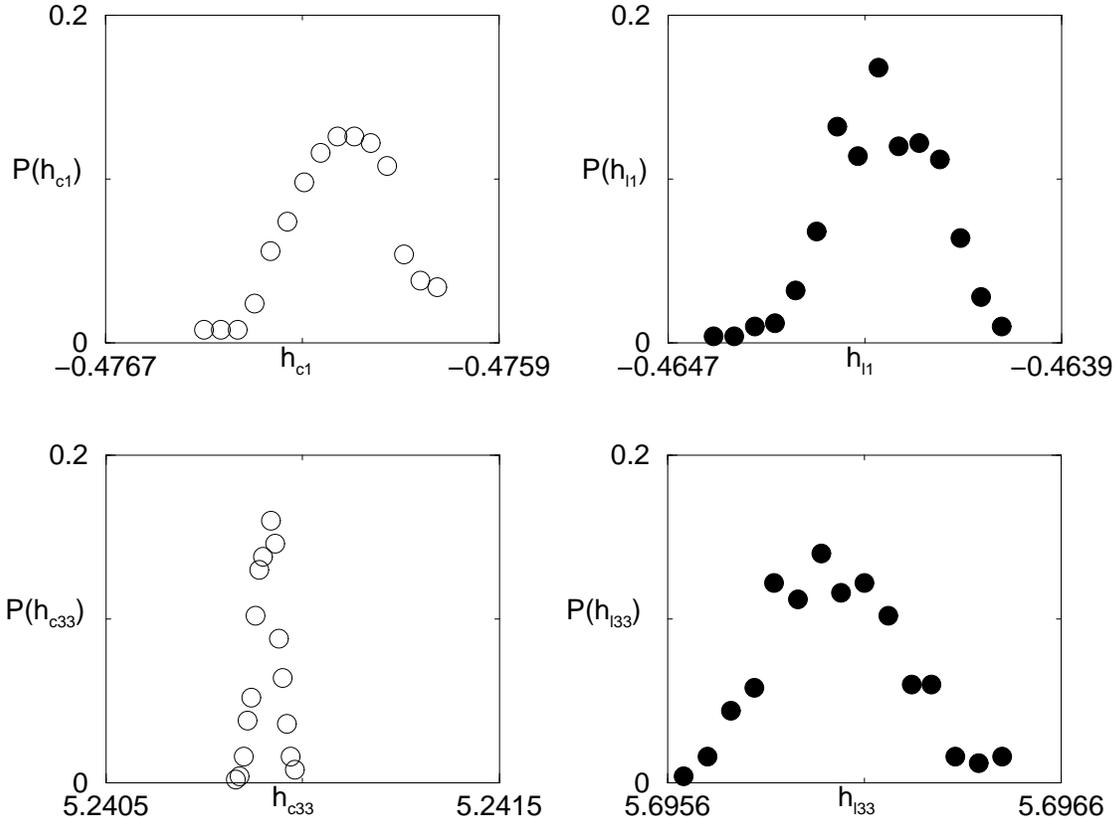,width=5.7in}}
\vspace{10pt}Ê\caption{The probability distribution 
of the cavity fields $P(h_c)$ (open circles) and 
of the local fields $P(h_l)$ (black circles) 
 for the species $1$ and $33$ after time $t = 100$.
 The system parameters
are the same of Fig.(1).}
	\label{P(hc)vsP(hl)}
	\end{figure}

\noindent
We have also analyzed the dynamics of the ecosystem when 
one species is absent. Specifically we considered the cavity
field, which is the field acting on the {\em i}-th population when this
population is absent \cite{MePaVir}.
In Fig.(2) we report the probability
distributions of the local and of the cavity fields obtained by our
simulations after a time $t = 100$ (expressed in arbitrary
units) in absence of
external noise and for two species (namely species $1$ and $33$).
We note that the probability distributions of the cavity fields differ
substantially from that of the local fields for the same species
unlike the spin glasses dynamics, where the two fields coincide.
We calculate also the same quantities in the presence of the external
noise. The results of our simulations are reported 
in Fig.s (3) and (4). The effect of the external noise is to overlap
the two fields in such a way that for some particular species
they coincide. Specifically this happens for the species $1$
(see Fig. (3)). For the species $33$ we obtain a partial 
overlap (see Fig.(4)). 

\begin{figure}[b!] 
\centerline{\epsfig{file=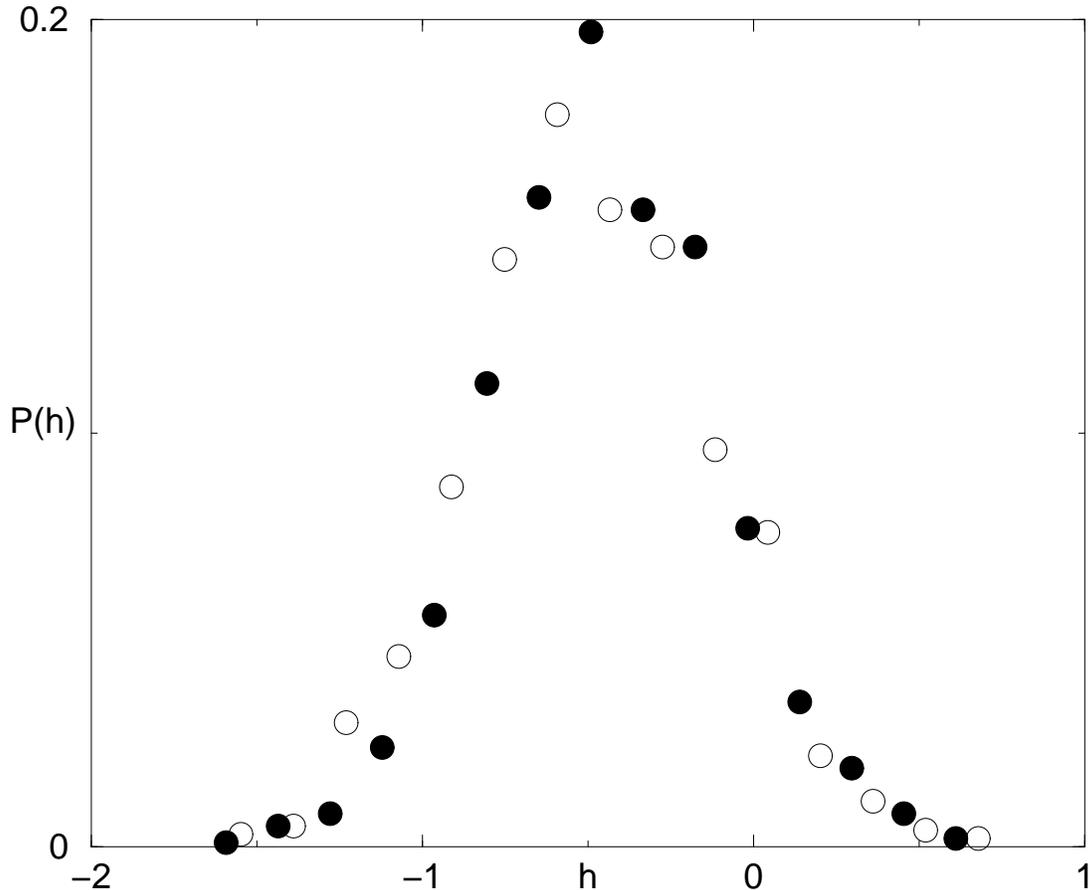,width=5.7in}}
\vspace{10pt}Ê\caption{The probability distribution $P(h_1)$
of the local (black circles) and of the cavity (open circles) 
fields for the species $1$ after time $t = 100$, in the 
presence of external noise. The noise intensity 
is $\epsilon = 0.1$.
 The other system parameters
are the same of Fig.(1).}
	\label{P(hl_hc_1)}
	\end{figure}

\noindent
We found this interesting phenomenon, which is reminiscent
of the phase transition phenomenon, for
some populations. The main reasons for this behaviour 
are: (i) all the populations are positive; (ii)
the particular structure of the attraction  basins of our ecosystem;
(iii) the initial conditions, which differ for the value of one
population, belong to
diffeent attraction basins.  
Some populations have a dynamical behaviour such 
that after a long time they influence in a significant way the
dynamics of other species. While in the presence of noise
all the populations seem to be equivalent from the
dynamical point of view.
We found also that for strong noise intensity
(namely $\epsilon =1$) all species extinguish on a 
long time scale ($t \approx 10^6$ a. u.). Whether extinction 
occurs for any value of noise intensity or not is still 
an open question, because of time-consuming 
numerical calculations.

\begin{figure}[b!] 
\centerline{\epsfig{file=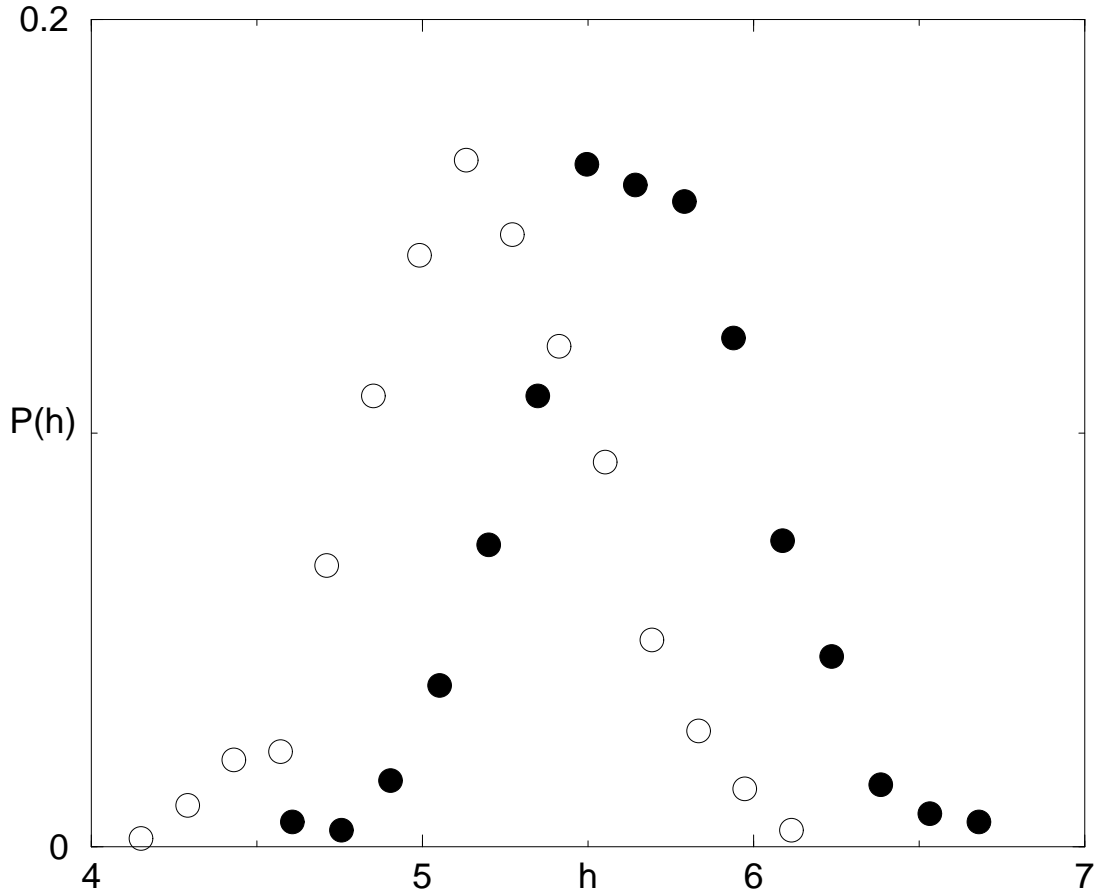,width=5.7in}}
\vspace{10pt}Ê\caption{The probability distribution  $P(h_{33})$
of the local (black circles) and of the cavity (open circles) 
fields for the species $33$ after time $t = 100$
in the 
presence of external noise. The noise intensity 
is $\epsilon = 0.1$.
 The other system parameters
are the same of Fig.(1).}
	\label{P(hl_hc_33)}
	\end{figure}

\section{CONCLUSIONS} 
\noindent 
We studied a stochastic model of an ecosystem of N 
interacting species. By means of an approximation of 
the integral equation, which gives the stochastic evolution 
of the system, we obtain analytical results reproducing 
very well almost all the transient. We investigate the role of the noise on
 the stability-instability transition and on the transient 
dynamics. For random interaction we obtain asymptotic behaviour for three
different nonlinear relaxation regimes. We obtain the 
stationary probability distribution of the population, which
is the sum of two contributions: (i) a delta function around $n = 0$
for the extinct species and (ii) a truncated Gaussian for the
alive species. When we switch on the external noise an interesting
phenomenon is observed: the local and the cavity fields, whose probability
distributions are different in the absence of noise, coincide for
some populations. This phenomenon can be ascribed to the peculiarity
of the attraction basins of our ecosystem. We have also
investigated the overlap between the asymptotic values of 
the populations
and the eigenvector of the interaction matrix with the maximum 
eigenvalue and we have not found any ordering regime 
phenomenon like in the spin
glasses system.  A more detailed
investigation concerning the probability distribution
of the populations and the local fields
in the presence of noise is the subject of
work in progress.

\section{ACKNOWLEDGMENTS} 
\noindent 
This work was supported in part by the National Institute of
Physics of Matter (INFM) and the Italian Ministry of
Scientific Research and University (MURST).

\end{document}